\documentclass[aps,prl,twocolumn,superscriptaddress,showpacs]{revtex4}
\usepackage{graphics}
\usepackage{subfigure}
\usepackage{longtable}
\usepackage{times}
\begin{document}

\preprint{}

\title{Strongly Correlated Electrons in the $\mathbf{\left[Ni(hmp)(ROH)X\right]_4}$ Single Molecule Magnet: A DFT+U Study}


\author{Chao Cao}
\affiliation{Department of Physics, University of Florida, Gainesville, FL 32611, U.S.A.}
\affiliation{Quantum Theory Project, University of Florida, Gainesville, FL 32611, U.S.A.}
\author{Stephen Hill}
\affiliation{Department of Physics, University of Florida, Gainesville, FL 32611, U.S.A.}
\author{Hai-Ping Cheng}
\affiliation{Department of Physics, University of Florida, Gainesville, FL 32611, U.S.A.}
\affiliation{Quantum Theory Project, University of Florida, Gainesville, FL 32611, U.S.A.}
%\email[Corresponding author, e-mail: ]{cheng@qtp.ufl.edu}
%\thanks{}


\date{\today}

\begin{abstract}
The single-molecule magnet $\mathrm{\left[Ni(hmp)(MeOH)Cl\right]_4}$
is studied using both density functional theory (DFT) and the DFT+U
method, and the results are compared. By incorporating a Hubbard-U
like term for both the nickel and oxygen atoms, the experimentally
determined ground state is successfully obtained, and the exchange
coupling constants derived from the DFT+U calculation agree with
experiment very well. The results show that the nickel 3d and oxygen
2p electrons in this molecule are strongly correlated, and thus the
inclusion of on-site Coulomb energies is crucial to obtaining the
correct results.
\end{abstract}

%\keywords{}

\maketitle

Single-molecule magnets (SMMs) have drawn much attention since their
discovery in 1991 \cite{SMM_GEN_0,SMM_GEN_1,SMM_GEN_2}. SMM crystals
contain ordered arrays of molecular nanomagnets, each possessing a
large spin ground state ($S=10$ for Mn$_{12}$-Ac) and a significant
uniaxial magneto-anisotropy ($DS_z^2$, with $D<0$). These two
ingredients give rise to a magnetic spectrum for an isolated
molecule in which the lowest lying levels correspond to the
`spin-up' and `spin-down' states ($m_s = \pm S$), separated by an
energy barrier of order $DS^2$. This barrier results in magnetic
bistability and hysteresis at low temperatures ($k_BT<<DS^2$). In
contrast to bulk ferromagnets, however, this hysteresis is intrinsic
to the individual molecules. There has, therefore, been much
interest in the potential implementation of SMMs as the elementary
memory units in both classical and quantum computers.

For most transition metal complexes (including Mn$_{12}$-Ac), the
intramolecular superexchange between the constituent ions is
predominantly antiferromagnetic (AFM). Nevertheless, due to spin
frustration effects, uncompensated moments of many tens of $\mu_B$
are often realized. However, the ability to engineer pure
ferromagnetic superexchange within a molecule is highly desirable
\cite{Harris}, because this ultimately removes one of the many
challenges in designing new and better SMMs. A relatively new series
which has attracted recent interest is
$\mathrm{\left[Ni(hmp)(ROH)Cl\right]_4}$ \cite{Ex_Bias_Ni_SMM,
EPR_Ni_SMM, SPP_SPIN_SMM, MAG_TUN_Ni_SMM, GSA_LIM_SMM} (hereon
denoted $\mathrm{Ni_4}$), where hmp is the anion of
2-hydroxymethylpyridine, and $\mathrm{R}$ is an alkyl substituent
such as methyl, ethyl, etc. Several experiments, including EPR
studies \cite{EPR_Ni_SMM} and magnetic susceptibility measurements
\cite{Ex_Bias_Ni_SMM} clearly show that the ground state of Ni$_4$
is ferromagnetic with total spin $S=4$. In this letter, we present
the results of detailed density functional theory (DFT) calculations
(including on-site Coulomb energies) which provide crucial insights
into the origin of this ferromagnetic state.

\begin{figure}[htp]
  \centering
  \subfigure[ ]{
    \scalebox{0.3}{\includegraphics{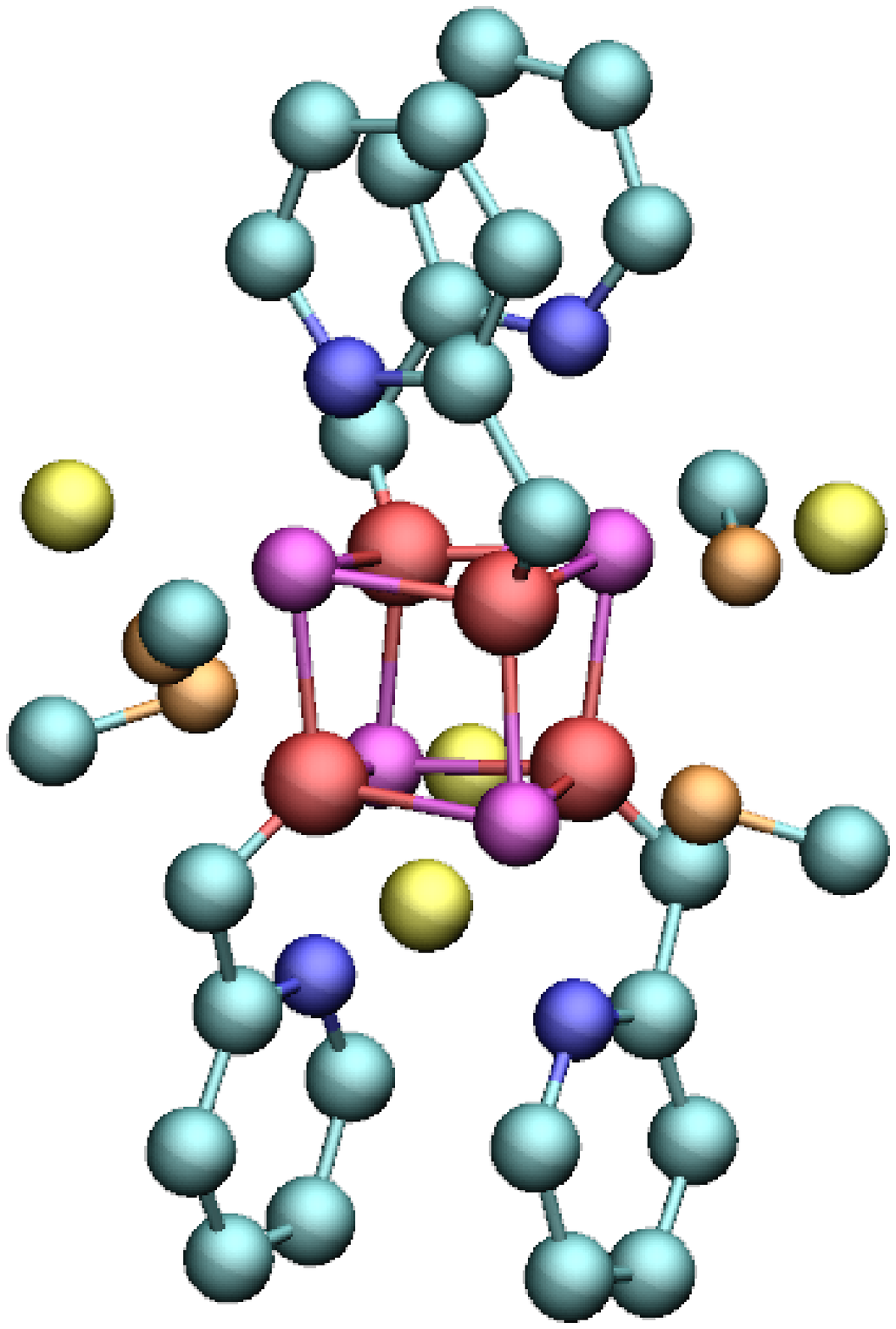}}
    \label{FIG_SMM_GEO}
  }
  \subfigure[ ]{
    \scalebox{0.5}{\includegraphics{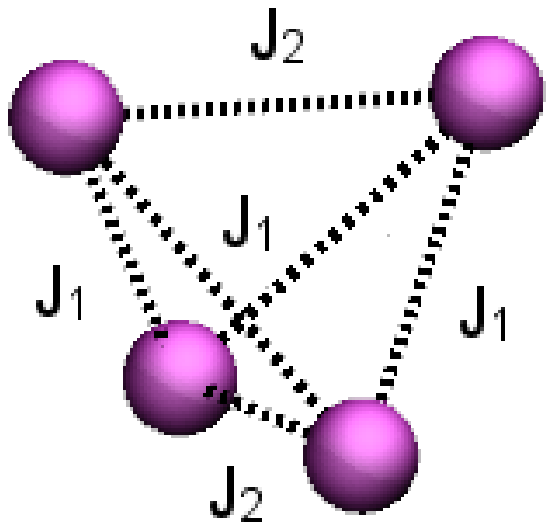}}
    \label{FIG_SMM_XCC}
  }
  \caption{Optimized structure (a) and exchange-coupling scheme (b) of $\mathrm{\left[Ni(hmp)(MeOH)Cl\right]_4}$. Magenta atoms are Ni; red atoms O(1); yellow atoms Cl; orange atoms other oxygens (i.e. O(2)); blue atoms N; and gray atoms C. The Ni and O(1) atoms form a slightly distorted cube.}
  \label{FIG_SMM_STRUCTURE}
\end{figure}

While DFT \cite{DFT_KS} has successfully explained the properties of
a variety of SMMs, including $\mathrm{Mn_{12}}$, $\mathrm{Mn_4}$,
$\mathrm{Co_4}$, $\mathrm{Fe_4}$
\cite{MAG_ANI_PEDERSON,Mn12_PARK,Mn4_PARK,Co4_BARUAH, Fe4_KORTUS},
and even some other nickel based SMMs \cite{Ni12_DFT}, it has so far
failed miserably for $\mathrm{Ni_4}$. Not only were the early
theoretical attempts unable to reproduce the correct ground state,
but the resulting coupling constants were also found to be
antiferromagnetic, and orders of magnitude higher than the
experimental values \cite{THEORY_Ni_SMM}. It was also found in the
calculation that the spin density is not quite localized around the
nickel atoms, as expected. Thus, it has been suggested that the
discrepancy between theory and experiments might arise due to the
small ``spin density leakage" in this system, resulting in spin
delocalization.

There is another possibility. Due to the localized nature of 3d
electrons, transition metal dioxides, including nickel oxides, are
known to be strongly correlated materials. The functioning core of
$\mathrm{\left[Ni(hmp)(ROH)Cl\right]_4}$, on the other hand, is a
cubic tetra-nickel oxide ($\mathrm{Ni_4O_4}$), which is structurally
very close to the nickel oxide complex. Therefore, it is more
probable that the lack of strong correlation in DFT is responsible
for this failure, and the ``spin density leakage" is just an
artifact. To justify this speculation, we calculated the electronic
structure of $\mathrm{\left[Ni(hmp)(MeOH)Cl\right]_4}$ using both
the DFT and DFT+U methods. The latter was introduced by V. I.
Anisimov et al. \cite{DFTU_ANISIMOV} and simplified by M. Cococcioni
et al. \cite{DFTU_MATTEO}.

All the reported calculations were done using the PWSCF package
\cite{PWSCF}, which utilizes PBE exchange-correlation functionals
\cite{DFT_PBEXC},  ultrasoft pseudopotentials \cite{VANDERBILT_PP},
and a plane-wave basis-set. We respectively chose energy cut-offs
for the wave functions and charge densities to be 40~Ry and 400~Ry
to ensure total energy convergence. The structure of the molecule
was optimized with a fixed total spin $S=4$ until the force on each
atom was smaller than 0.01 eV/$\mathrm{\AA}$. The relaxed structure
is in good agreement with experimental results. The same structure
was then used for some of the AFM states ($S=0$) and $S=2$, as well
as for all of the DFT+U calculations. Due to symmetry restrictions,
we only simulated the $S=4$, $S=2$ and AFM ($S=0$) states. For the
DFT+U calculations, a self-consistent Hubbard-U method
\cite{DFTU_MATTEO} has been incorporated to determine the U value
for Ni, which turns out to be 6.20~eV for this system. For oxygen,
we took the well-established U value of 5.90~eV \cite{DFTUPD}. For
the DOS and projected DOS, we used 0.1~eV gaussian smearing to
smooth the results.
%consulted several previous papers, and took the well-established
%Hubbard-U parameter for Ni (4.58 eV) \cite{DFTU_MATTEO} and O (5.90
%eV) \cite{DFTUPD}. For the DOS and projected DOS, we used 0.1 eV
%gaussian smearing to smooth the results.

\begin{figure}[htp]
  \centering
  \scalebox{0.60}{\rotatebox{270}{\includegraphics{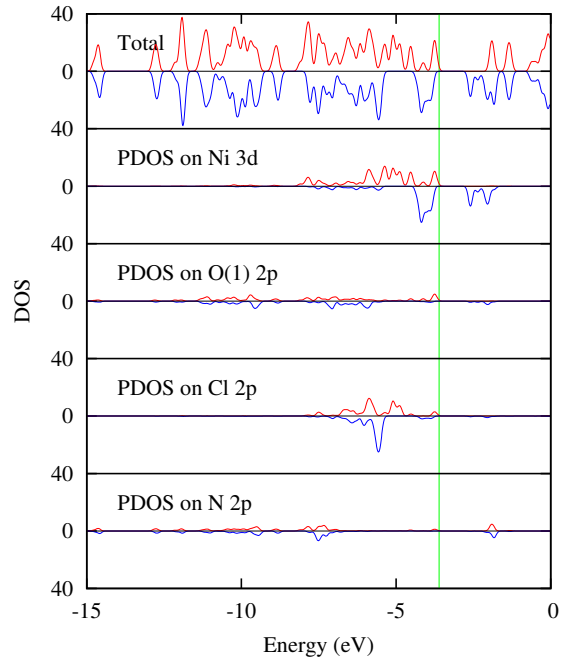}}}
  \caption{Total DOS and PDOS on the 2p and 3d orbitals of $\mathrm{\left[Ni(hmp)(MeOH)Cl\right]_4}$ for the $S=4$ state using $\mathrm{DFT}$. Red lines represent $\alpha$-spin and blue lines represent $\beta$-spin. The green line is the Fermi level, $E_F$. The DOS were drawn to the same scale for comparison purposes.}
  \label{FIG_PDOS_DFT}
\end{figure}

The DFT calculations confirmed Park et al.'s findings. The optimized
structure is shown in Figure \ref{FIG_SMM_GEO}. In order to better
display the geometry, we hide all hydrogen atoms. The four nickel
atoms and four oxygen atoms on the hmp group (we call them O(1) from
now on) define a slightly distorted cube. The AFM state ($S=0$)
turns out to be the ground state, which is 14.8~meV lower than the
$S=2$ state and 35.1~meV lower than the $S=4$ state (table
\ref{TAB_ETOT}, column DFT). Using L$\mathrm{\ddot{o}}$wdin's charge
analysis, one can see that about a 1.48--1.50~$\mu_B$ magnetic
moment is found on each nickel atom, and each of the four O(1) atoms
contributes a 0.1--0.26~$\mu_B$ magnetic moment (table
\ref{TAB_MAGMOM}, column DFT). The Heisenberg Hamiltonian in general
can be written as:
\begin{equation}
H_{ex}=\sum_{i<j}J_{ij}\mathbf{S}_i\cdot\mathbf{S}_j.
\end{equation}
Considering the molecule's $S_4$ symmetry (Figure
\ref{FIG_SMM_XCC}), the six $J_{ij}$s reduced to just two values,
i.e.:
\begin{eqnarray}
H_{ex}=J_1(\mathbf{S}_1\cdot\mathbf{S}_2+\mathbf{S}_2\cdot\mathbf{S}_3+\mathbf{S}_3\cdot\mathbf{S}_4+\mathbf{S}_4\cdot\mathbf{S}_1)
\cr
+J_2(\mathbf{S}_1\cdot\mathbf{S}_3+\mathbf{S}_2\cdot\mathbf{S}_4).
  \label{EQ_HSB_HAM}
\end{eqnarray}
We then determined these exchange coupling constants by fitting the
expression of Eq. (2) to the obtained energies of the $S=4,2$ and 0
spin states, giving $J_1=3.54$~meV and $J_2=2.12$~meV. The positive
sign for both numbers indicates AFM coupling. From the total
electronic density of states (DOS), as well as the projected density
of states (PDOS) onto Ni (Figure \ref{FIG_PDOS_DFT}), it is clear
that, in the DFT calculation, the nickel atom 3d orbitals dominate
both the highest occupied (HOMO) and lowest unoccupied molecular
orbitals (LUMO), so that the HOMO-LUMO gap is of the d-d type.

The DFT+U calculations were performed in two stages. In the first
stage (we call this $\mathrm{DFT+U^d}$), we turned on the Hubbard-U
parameter for the nickel 3d orbitals only, like most current DFT+U
calculations; in the second stage (we call this
$\mathrm{DFT+U^{p+d}}$), we turned on the U parameter for both the
nickel 3d and O(1) 2p orbitals. In fact, it is known that Coulomb
interactions between oxygen 2p electrons are comparable to those
between d electrons \cite{PRB5,PRB6}, and should hence be taken into
consideration as well. However, since oxygen usually bares a fully
occupied p-shell, this correlation effect is often thought to be
neglegible. Therefore, in most cases, $\mathrm{DFT+U^{d}}$ can
already yield a satisfactory description of the ground-state without
oxygen 2p-electron corrections. Nevertheless, DFT+U has to be taken
into consideration explicitly here for both the 3d and oxygen 2p
electrons in order to obtain the correct ground state for this
molecule.

\begin{table}
\begin{ruledtabular}
\begin{tabular}{c|ccc}
  &DFT&$\mathrm{DFT+U^{d}}$&$\mathrm{DFT+U^{p+d}}$\\
  \hline
  AFM (S=0)&0.0000&0.00000&0.000000\\
  S=2&0.0011&0.00012&$-0.000069$\\
  S=4&0.0026&0.00019&$-0.000368$\\
\end{tabular}
\end{ruledtabular}
\caption{Total energies in Rydbergs. All numbers are relative to the
AFM state ($S=0$)} \label{TAB_ETOT}
\end{table}

\begin{table}
\begin{ruledtabular}
\begin{tabular}{c|ccc|ccc|ccc}
  &\multicolumn{3}{c}{DFT}&\multicolumn{3}{c}{$\mathrm{DFT+U^{d}}$}&\multicolumn{3}{c}{$\mathrm{DFT+U^{p+d}}$}\\
  \hline
  &AFM&S=2&S=4&AFM&S=2&S=4&AFM&S=2&S=4\\
  \hline
Ni  & 1.49& 1.49& 1.50& 1.68& 1.68& 1.68& 1.74& 1.74& 1.74\\
O(1)& 0.11& 0.13& 0.26& 0.06& 0.08& 0.16& 0.04& 0.06& 0.11\\
Cl  & 0.09& 0.09& 0.09& 0.06& 0.06& 0.06& 0.06& 0.06& 0.06\\
N   & 0.08& 0.08& 0.08& 0.06& 0.05& 0.05& 0.05& 0.05& 0.05\\
O(2)& 0.06& 0.06& 0.06& 0.03& 0.03& 0.03& 0.03& 0.03& 0.03\\
\end{tabular}
\end{ruledtabular}
\caption{Magnetic moments (in $\mu_B$) captured by Ni, O(1), Cl, N
and O(2) atoms. AFM indicates the antiferromagnetic state ($S=0$).
All numbers are averaged over the same species.}
%\caption{Magnetic moments (in $\mu_B$) captured by each Ni and O(1) atom. AFM indicates the antiferromagnetic state (S=0)}
\label{TAB_MAGMOM}
\end{table}

\begin{figure}[htp]
  \centering
  \scalebox{0.60}{\rotatebox{270}{\includegraphics{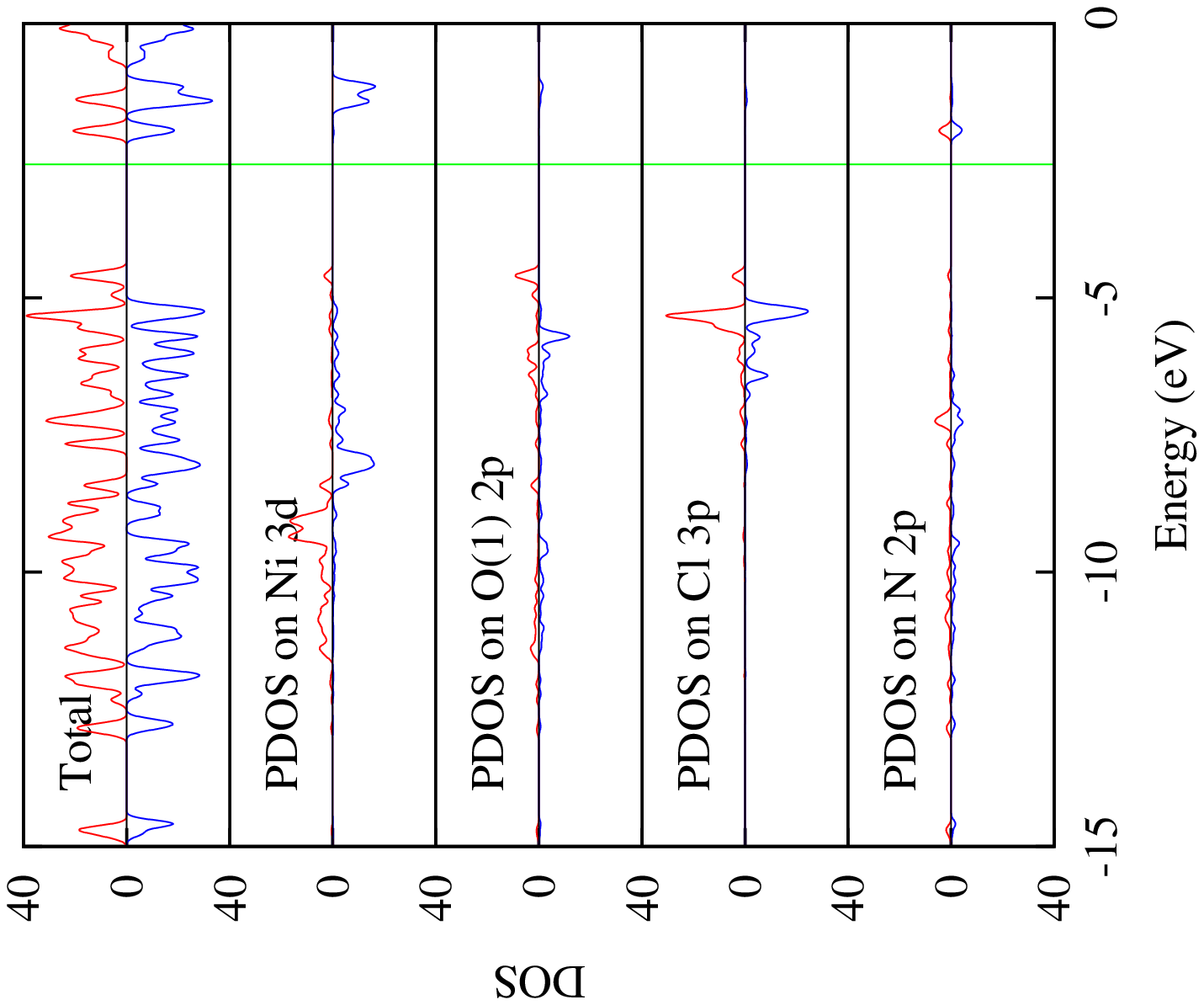}}}
  \caption{Total DOS and PDOS on the 2p and 3d orbitals of $\mathrm{\left[Ni(hmp)(MeOH)Cl\right]_4}$ for the $S=4$ state using $\mathrm{DFT+U^{d}}$. Red lines represent $\alpha$-spin and blue lines represent $\beta$-spin. The green line is the Fermi level, $E_F$. DOS were drawn to the same scale for comparison purposes.}
  \label{FIG_PDOS_DFTUD}
\end{figure}

\begin{figure}[htp]
  \centering
  \scalebox{0.60}{\rotatebox{270}{\includegraphics{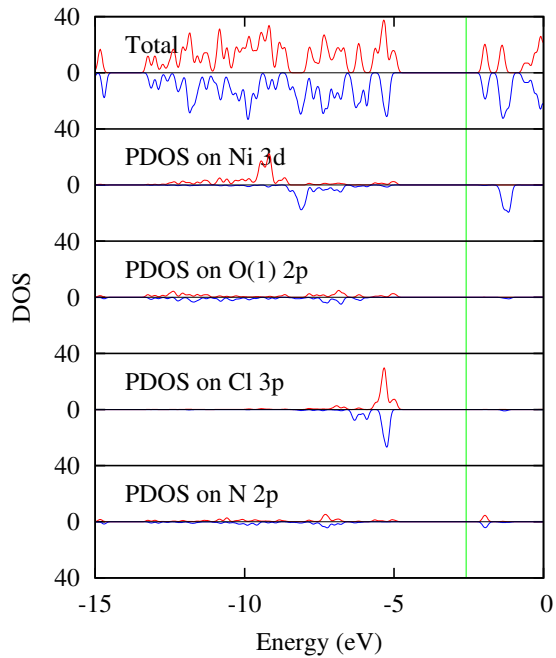}}}
  \caption{Total DOS and PDOS on the 2p and 3d orbitals of $\mathrm{\left[Ni(hmp)(MeOH)Cl\right]_4}$ for the $S=4$ ground-state using $\mathrm{DFT+U^{p+d}}$. Red lines represent $\alpha$-spin and blue lines represent $\beta$-spin. The green line is the Fermi level, $E_F$. DOS were drawn to the same scale for comparison purposes.}
  \label{FIG_PDOS_DFTUPD}
\end{figure}

The DFT+U energies are shown in table \ref{TAB_ETOT}. By turning on
DFT+U for the nickel atoms only, the energy differences between
different spin states were greatly reduced, hence giving much
smaller numbers for the exchange coupling constants. However, the
ground state here is still AFM ($S=0$), and the energies for the
$S=4$ and $S=2$ states relative to the AFM state are  2.61~meV and
1.60~meV, respectively. But once we take into consideration the
strong coulomb interactions for both the Ni and O(1) atoms, the
order is reversed, yielding correctly a $S=4$ ground state. A
spin-unrestricted calculation also confirmed this discovery. The
$S=2$ state is now 0.94~meV lower than the AFM state, and the $S=4$
ground state is 5.00~meV lower. Using these values, we obtained
ferromagnetic exchange-coupling constants for the
$\mathrm{DFT+U^{p+d}}$ calculation from a fit to equation 2, i.e.
$J_1=-0.50$~meV and $J_2=-0.68$ meV \ref{EQ_HSB_HAM}. These results
match experiment reasonably well ($-0.68$ and $-2.28$~meV)
\cite{Yang_INORG_CHEM}.

\begin{figure}[htp]
  \centering
  \scalebox{0.55}{\rotatebox{270}{\includegraphics{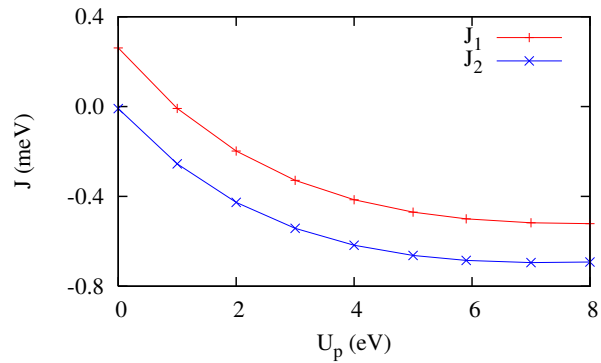}}}
  \caption{Variation of $J_1$ and $J_2$ with respect to different U$_p$ for O(1).}
  \label{FIG_UJDEP}
\end{figure}
To better understand the contribution of the Hubbard-U like term, we
first performed L$\mathrm{\ddot{o}}$wdin's charge analysis to
calculate the magnetic moments captured by the Ni, O(1), Cl and N and
O(2) atoms (table \ref{TAB_MAGMOM}). The results show that the spin
density is more localized in the $\mathrm{DFT+U^{p+d}}$ calculations
(1.74~$\mu_B$) than in normal DFT (around 1.50~$\mu_B$), and that
the magnetic moments found on the O(1) atoms are greatly reduced.
This is because the strong on-site Coulomb interaction prevents the
hybridization between the nickel 3d and oxygen 2p orbitals, thus
preventing the unphysical ``spin-leakage". The DFT calculations
favor the AFM ground state because the lack of on-site energy tends
to couple electrons with opposite spin projections, and thus lead to
the incorrect ground state. The local magnetic moment on an
individual atom is a measurable quantity using NMR, and thus can be
used to validate these theoretical predictions.

Total DOS and PDOS in the $\mathrm{DFT+U^{d}}$ method (Figure
\ref{FIG_PDOS_DFTUD}) and $\mathrm{DFT+U^{p+d}}$ method (Figure
\ref{FIG_PDOS_DFTUPD}) were also calculated \cite{EPAPS}. In contrast to the DFT
results, the dominant contribution to the HOMO and LUMO states in
both DFT+U calculations is now from the 2p (Cl and O) and 3d (Ni)
orbitals, respectively (Figure \ref{FIG_PDOS_DFTUPD}). We do not
have direct experimental results to compare with this feature,
however, for nickel oxide it is well known that DFT gives incorrect
PDOS contributions. Early experiments and calculations
\cite{GAP_OPT_NiO, GAP_LSDAU, XPS_NiO, LDAU_APW, LDAU_PAW} show that
instead of a d-d gap given by DFT, nickel oxides actually have a p-d
gap. The largest spin density contribution, of course, is still from
the Ni 3d electrons. For the $S=4$ ground state, the
$\mathrm{DFT+U^{p+d}}$ calculation yields a LUMO-HOMO gap of
2.95~eV, which is from majority spin to minority spin; in the
$\mathrm{DFT+U^{d}}$ and DFT calculations, these numbers are 2.56~eV
and 1.09~eV, respectively. Since DFT has been known to underestimate
energy gaps and excitation states, DFT+U calculations have proven
necessary in order to obtain agreement with experiments such as
resonant inelastic X-ray scattering (RIXS) and XPS \cite{Mn12_LDAU}.
The present study also likely calls into question recent reports of
HOMO-LUMO gaps of fractions of an eV in the Fe$_8$ SMM
\cite{Baruah}.

Finally, we have also repeated $\mathrm{DFT+U^d}$ and
$\mathrm{DFT+U^{p+d}}$ calculations with $U^d$ for nickel ranging
from 4.58 to 6.20 eV and $U^p$ for oxygen ranging from 1.0 to 8.0
eV. The energetic order of spin states remains the same, and the
magnitudes of the energy differences between spin states are
more-or-less insensitive to variations of $U^p$ from 3.0 to 8.0 eV,
as seen in Figure~\ref{FIG_UJDEP}. This clearly demonstrates the
reliability and robustness of our results.

In conclusion, we have performed DFT and DFT+U calculations for
$\mathrm{\left[Ni(hmp)(MeOH)Cl\right]_4}$. Because of the strong
correlation effects in this system, the DFT calculation fails due to
the fact that the lack of on-site energy unphysically encourages the
hybridization of orbitals, leading to AFM coupling. The inclusion of
a Hubbard-U like term for both the Ni 3d and O(1) 2p electrons
greatly enhances the localization for both states, and is essential
in order to obtain the correct ferromagnetic ground state and
exchange-coupling constants. After taking both corrections into
consideration, these properties were successfully reproduced by the
calculations. We then analyzed the DOS and projected DOS of the
system, and the calculation predicts that the optical transition
from HOMO to LUMO is p-d like, and the gap is 2.95~eV.

\begin{acknowledgments}
This work is supported by DOE DE-FG02-02ER45995 (H.-P. Cheng and C. Cao),
NSF/DMR/ITR-0218957 (H.-P. Cheng and C. Cao), NSF DMR0239481 (S. Hill),
 and NSF DMR0506946 (S. Hill). The authors want to thank NERSC,
CNMS/ORNL and the University of Florida High Performance Computing
Center for providing computational resources and support that have
contributed to the research results reported within this paper.
\end{acknowledgments}

\bibliographystyle{apsrev}
\bibliography{PRL}

\end{document}